\documentclass{article}
\usepackage{amssymb,amsmath}

\usepackage{sfmath}
\usepackage{hyperref}

\newcommand{\rmssf}[1]{\relax\ifmmode\mathsf{#1}\else\textsf{#1}\fi}

\begin{document}

\title{An OpenMath Content Dictionary for\\
 Tensor Concepts}
\author{Joseph B. Collins\\
Naval Research Laboratory\\
4555 Overlook Ave, SW\\
 Washington, DC  20375-5337}

\date{}

\maketitle

\begin{abstract}
We introduce a new OpenMath content dictionary named ``tensor1''
containing symbols for the expression of tensor formulas.
These symbols support the expression of non-Cartesian coordinates and
invariant, multilinear expressions in the context of coordinate
transformations.
While current OpenMath symbols support the expression of linear
algebra formulas using matrices and vectors, we find that there is an
underlying assumption of Cartesian, or {\em standard}, coordinates
that makes the expression of general tensor formulas difficult, if not
impossible.
In introducing these new OpenMath symbols for the expression of tensor
formulas, we attempt to maintain, as much as possible, consistency
with prior OpenMath symbol definitions for linear algebra.
\footnote{The final publication of this paper is
 available at www.springerlink.com}

\end{abstract}

\section{Introduction}

In scientific and engineering disciplines there are many uses of
tensor notation.
A principal reason for the need for tensors is that the laws of
physics are best formulated as tensor equations.
Tensor equations are used for two reasons: first, the physical laws of
greatest interest are those that may be stated in a form that is
independent of the choice of coordinates, and secondly; expressing the
laws of physics differently for each choice of coordinates becomes
cumbersome to maintain.
While from a theoretical perspective it is desirable to be able to
express the laws of physics in a form that is independent of
coordinate frame, the application of those laws to the prediction of
the dynamics of physical objects requires that we do ultimately specify
the values in some coordinate frame.
Much of this discussion might be moot were scientists and engineers to
confine themselves to one frame, e.g., Cartesian coordinates, but such
is not the case.
Non-Cartesian coordinates are useful for curved geometries and
because closed form solutions to applied models in classical physics,
which rely on the separation of variables method of solving partial
differential equations, often exist in them.
For example, the Laplace equation is separable in thirteen
coordinate systems \cite{MorseFeshbach}.
One may also take as a definition of need that these concepts are
included in the ISO standards defining the necessary mathematical
symbols in the International System (SI) of Quantities and Units
\cite{siMath}.

While we specify the new OpenMath symbols for tensor concepts, we
attempt to maintain consistency with pre-existing OpenMath symbols
\cite{OpenMath}.
The symbols within OpenMath content dictionaries support the
expression of a wealth of mathematical concepts.
Determining whether or not additional symbols are needed requires
consideration based upon both necessity and convenience.
Advancing new symbols using arguments based upon mathematical
necessity only implies that a proof is at hand showing that a
particular concept cannot be expressed using existing OpenMath
symbols.
Since such proofs would be difficult, if not impossible, in practice,
convincing arguments for new symbols are more likely to be made based
on a combination of practical economy, practical necessity, and
convenience: this is certainly the case here.
The symbols we introduce are motivated by the need for easily and
directly capturing the relevant semantics in the expression of tensor
formulas.

OpenMath symbols exist for the specification of matrices and
vectors.
These are documented in the content dictionaries linalg1, 
linalg2, linalg3, linalg4, linalg5, and two dictionaries named
linalg6. 
Within these dictionaries there are two representations: one for row
vectors and one for column vectors, with the row representation being
labeled ``official'' in preference to the column representation.
In the row vector representation a matrix is a column of rows, and in the
column vector representation it is a row of columns.
We find that the two representations, i.e., the row representation of
vectors and the column representation of vectors, appear to be
alternative, equivalent representations, related by a transpose
operation, rather than dual representations, such as vectors and
covectors where row and column representations of a vector are related
via a general metric tensor.
We also find, from the few examples given and from the general lack of
reference to bases, that the row and column vector semantics appear to
assume use of the {\em standard}, or Cartesian, basis only, and, in
particular, with a simple Euclidean metric.
For example, the scalar product is given as 
${\bf u} \cdot {\bf v}= \sum_i u_i v_i$.
In the row representation, the vector components resulting from a
matrix-vector multiplication appear as the results of scalar products
between matrix rows and a row vector, i.e., a scalar product takes as
its arguments two vectors from the same vector space.
Given these observations, it is not clear that in using these existing
representations it is easy, or even possible, to express the
semantics of tensors as they are typically used by engineers and
scientists.
For these reasons we introduce symbols that are expressly to be used
for specifying tensor formulas.

\section{Tensor Review}

To motivate our choice of OpenMath symbols for specifying tensors, we
briefly review some tensor basics.
By definition, a tensor is a multilinear mapping that maps vectors and
covectors to a scalar field.
A tensor is itself an element of the space defined by a tensor
product of covector and vector spaces.
Among scientists and engineers, tensor formulas are commonly
written using their indexed {\em components}.

\subsection{Coordinate Frames}

To begin the discussion, we note that an arbitrary point in
$n$-dimensional space, ${\bf R}^n$, is typically specified by its $n$
Cartesian, or {\em standard}, coordinates, $x^i$.
The point's {\em position vector}, is then written as
${\bf r} = \sum_i x^i {\bf e}_i$, where the ${\bf e}_i$ are
orthonormal Cartesian basis vectors and are
constant, i.e., not a function of the coordinates, for a given
Cartesian frame.
Using the original Cartesian frame, alternative coordinates
may be defined as functions of the Cartesian coordinates in the
original frame, e.g., ${x'}^i= {x'}^i (x^1, ..., x^n)$, which may be
nonlinear in the $x^i$.

Spatial coordinates are sometimes expressed 
as indexed quantities, such as $(x^1, x^2, x^3)$,
or
having individual names, such as $(x, y, z)$.
In presentation, different kinds of indexes may appear much the same,
but in content markup we must be more discriminating.
For example, vector components are one kind of indexed quantity.
It would be a mistake, however, to consider the tuple of
spatial coordinates to be a vector in the general case.
While it may seem to be a vector in Cartesian coordinates, i.e., a
position vector, this is not the case in, for example, polar coordinates.
For general coordinates the vector addition of coordinate position
tuples does not appear to have a defined meaning, i.e., the meaning of

coordinates(Point1) + coordinates(Point2) = coordinates(Point3) \\
is not preserved under general cordinate transformation.

Considering this, the most we should say is that the variables describing
the coordinates of an arbitrary point in a space comprise an $n$-tuple.
This appears to be similar to the notion of an OpenMath {\em context}
\cite{KohlhaseRabe}, i.e., an $n$-tuple of variables.
Consequently, while we need to represent $x^i$, it is inappropriate to
do this using the vector\_selector symbol, the vector component
accessor defined in the OpenMath linalg1 content dictionary.
For this reason we introduce the {\em tuple} and {\em tuple\_selector}
symbols.
The symbol, tuple, is an n-ary function that returns an $n$-tuple of
its arguments in the order that they are presented.
The symbol, tuple\_selector, takes two arguments, an $n$-tuple, and an
index, a natural number less than or equal to $n$, and returns the
indexed element.

Since the Cartesian frame is most often used, including in the
definition of coordinate transformations and the definition of
non-Cartesian frames, we find it useful to have symbols to express the
base concepts of Cartesian coordinates.
We propose the symbol {\em Cartesian} which takes a single argument, a
natural number, and returns the Cartesian coordinate, of a
right-handed Cartesian frame, corresponding to the value of the
argument.
The standard representation of Cartesian 3-space may then be
represented by either

tuple(x, y, z) = tuple(Cartesian(1), Cartesian(2), Cartesian(3))\\
or as

tuple\_selector(i, {\bf x}) = Cartesian(i).\\
Coordinate transformations may then be defined as functions on the
Cartesian coordinates.

The full meaning of the Cartesian coordinate variables comes from
their combination with the basis vectors for the Cartesian frame.
The commonly used orthonormal basis vectors for the Cartesian frame
are given by the symbol {\em unit\_Cartesian}, i.e., unit\_Cartesian
takes a single natural number as its argument and returns the
corresponding unit vector, say, ${\bf e}_i$.
Other representations are easily assigned, such as

tuple($\hat i$, $\hat j$, $\hat k$) =
tuple(unit\_Cartesian(1), unit\_Cartesian(2), unit\_Cartesian(3)).\\
Basis vectors, ${\bf g}_i$, for transformed coordinates, ${x'}^i$, are given by

${\bf g}_i = \sum_j{{\partial x^j} \over {\partial {x'}^i}} {\bf e}_j$.

\subsection{Vectors and Covectors}

To describe tensors we must also give prior description to vectors and
covectors.
A vector, ${\bf v}$, may be specified by components $v^i$ with respect to
an arbitrary, ordered, vector space basis,
$\left( {\bf g}_1, ..., {\bf g}_n \right)$, as ${\bf v} = \sum_i v^i {\bf g}_i$.
These basis vectors, ${\bf g}_i$, are generally the tangent vectors
with respect to the spatial coordinates, e.g., ${x}^i$.
In curvilinear coordinates these general basis vectors are clearly
functions of the coordinates.
A dual, covector space may be defined relative to a given vector
space.
A dual space is defined as a set of linear functionals on the vector
space and is spanned by a set of basis elements,
$\left( {\bf g}^1, ..., {\bf g}^n \right)$,
such that ${\bf g}^i\left({\bf g}_j\right) = \delta^i_{\; j}$,
where $\delta^i_{\; j}$ is the Kronecker tensor.
The symbol, {\em Kronecker\_tensor}, has components,
$\delta^i_{\; j}$,
equal to one when $i = j$ and zero otherwise.

The presentation of the indexes, either raised or lowered, on basis
vectors, basis covectors, vector components, or covector components,
generally indicates how the components transform.
With a transformation of coordinates, indexed tensor quantities
transform either {\em covariantly}, as do the basis vectors, ${\bf g}_i$, or
they transform {\em contravariantly}, as do vector components, $v^i$,
or, for example, coordinate differentials, $d{x}^i$.
Transforming from coordinates $x^i$ to coordinates ${x'}^i$, the
covariant transformation is defined by the transformation of the basis
vectors:

${\bf g'}_i = \sum_k{{\partial x^k} \over {\partial {x'}^i}} {\bf g}_k$.\\
The contravariant transformation of a vector's components is given
by:

${v'}^l = \sum_k{{\partial {x'}^l} \over {\partial x^k}} {v}^k$.\\
The covariant transformation of the components of a covector, u, is given by

${u'}_j = \sum_k{{\partial x^k} \over {\partial {x'}^j}} u_k$.

Vectors whose components transform contravariantly, and their
covectors, whose components transform covariantly, are tensors.
Many, but not all, vector quantities are tensors.
For example, the coordinates themselves, $x^i$, are referred
to as the components of a {\em position vector} (in Cartesian
coordinates), ${\bf x}$  or ${\bf r}$, which is {\em not} a tensor.
(We have already noted that the position vector in Cartesian
coordinates, defined as a tuple of position coordinates, does not
generally preserve its meaning after coordinate transformation).

In general, tensors may be created by tensor (outer) products of vectors
and covectors, contracted products of tensors, and sums of tensors of
the same order.
Order one tensors are contravariant or covariant vectors, while
order zero tensors are scalars.
The order of a higher order tensor is just the necessary number of
vectors and covectors multiplied together, using the tensor product,
to create it.

For the purpose of describing tensor formulas in content markup, we
introduce the OpenMath symbols
{\em tensor\_selector}, {\em contra\_index}, and {\em covar\_index},
which are applied to a natural number, returning the appropriate index.
In standard tensor notation, a contravariant index is represented as a
superscripted index and a covariant index is represented as a
subscripted index.
The contra\_index and covar\_index symbols are so named because
characterizing the indexes of tensor quantities as being contravariant
or covariant captures the semantics.

In engineering and scientific applications standard matrix-vector
multiplication is consistent with tensor notation when interpreted as
a matrix multiplying a column vector from the left, resulting in a
column vector.
Each of the components of the result are arrived at by applying the
rows of the matrix to the column vector being multiplied.
It is consistent with this common usage to identify the components of
a column vector using the contravariant, superscripted index as 
a row index, and to identify the components of a row vector using the
covariant, subscripted index as a column index.
The matrix-vector multiplication is then represented as
$u^i = \sum_j M^i_{\; j} v^j$.
It is common in tensor notation to suppress the explicit summation in
such an expression using the Einstein Summation Convention.

While it is common to implicitly assume the use of standard, or
Cartesian coordinates, in which case the distinction between
superscripts and subscripts appears superfluous, this is not so with
tensor notation: a vector may be specified by its components relative
to some general, non-Cartesian basis.
As pointed out above, the basis vectors of an arbitrary, ordered basis
of a vector space transform covariantly, hence they are indexed
using the symbol, covar\_index.
Similarly, the basis covectors, derived from the same arbitrary, ordered
basis of the vector space, transform contravariantly, hence they are
indexed using the symbol, contra\_index.

We introduce, then, the basis\_selector symbol as a binary
operator, taking as its arguments:

1) an ordered basis, a tuple of linearly independent vectors that
spans some vector space;

2) either a covar\_index or contra\_index symbol applied to a natural
number.\\
The basis\_selector operator returns a basis vector of the vector
space when a covar\_index symbol is passed and returns a basis
covector of the dual vector space when a contra\_index is passed.

\subsection{Higher Order Tensors}

To write expressions using tensor components, we use the symbol,
tensor\_selector.
The tensor\_selector symbol returns a scalar and takes three arguments:

1) a tensor;

2) a tuple of contra\_index and covar\_index symbols, and, finally;

3) a frame, an ordered set of basis vectors.

The sum total of indexes used, both contra\_index's and
covar\_index's, must be the same as the order of the tensor.
The contravariant and covariant indexes, taken together, are totally
ordered, and refer to a matrix of tensor components, which are assumed
to be in 'row-major' order, regardless of whether the indexes are
contra\_index's or covar\_index's.
By use of the term row-major order, we do not attribute any special
meaning to whether an index is considered a row index or a column
index, rather we merely mean that for the serial traversal of an arbitrarily
dimensioned array used to store an indexed quantity, the rightmost index
varies fastest.
The assumption of this convention allows the unambiguous assignment of
indexed matrix component values to indexed tensor components.

The scalar returned by tensor\_selector is the tensor component.
For example, the contravariant components of a vector are identified
by applying tensor\_selector to the vector and a contra\_index.
Components of higher order tensors are identified  by use of multiple
contra\_index and covar\_index symbols.
The final argument, the frame, is necessary when one needs to specify
a tensor expression that is dependent on the basis or on multiple
bases, as in a transformation expression.
As tensor formulas are commonly made without regard to basis, often no
basis is required, and so any single, consistent basis is sufficient
in this case, such as Cartesian.
It is also suggested that a special value, called ``unspecified'',
might be used.

A general tensor is usually indicated with a capital letter.
Its coordinates may be represented using a sequence of contra\_index
and / or covar\_index symbols.
The tensor itself may be represented by taking the product

${\bf T}=\sum_{ij}T^{ij}{\bf g}_i{\bf g}_j$.\\
The Einstein summation convention is normally implicitly applied to
the product of two tensors whose components are represented with
matching contravariant and covariant indexes.
In content markup this summation should be explicit since there is
otherwise no content markup to indicate the fact that these indexes
are bound variables.

Finally, we define a couple more symbols for
specific tensor and vector quantities.
First, there is the {\em metric\_tensor} which defines the geometric
features of the vector space, such as length.
Its components are represented as $g_{ij}$, a symmetric,
non-degenerate, covariant, bilinear form defined by
 $\left( ds
\right)^2 = g_{ij} d{x'}^id{x'}^j$, where $ds$ is the differential length
element and $d{x'}^i$ are the differential changes in spatial
coordinates.
This is a generalization of the simple Euclidean metric given by the
scalar product.
Covariant components and contravariant components of a vector, or row
and column representations of a vector, are related by
$v_i = \sum_j g_{ij}v^j$ and squared length, or squared norm, of a vector is
$\left| {\bf v} \right|^2 = \sum_{ij}g_{ij}v^iv^j = \sum_jv_jv^j$.

Lastly, we define the {\em Levi-Civita} symbol, the so-called
permutation tensor. 
It takes one argument, the dimension of the space.
Its components may be indexed with the contra\_index and / or
covar\_index symbols.

\subsection{Conclusion}

We have introduced a number of OpenMath symbols for the expression of
tensor formulas.
They are tuple, tuple\_selector, Cartesian, unit\_Cartesian,
Kronecker\_tensor, basis\_selector, tensor\_selector, contra\_index,
covar\_index, metric\_tensor, and Levi-Civita.
Using the  tuple, tuple\_selector, Cartesian, and unit\_Cartesian
symbols we can build finite dimensioned Cartesian frames and define
differentiable coordinate transformations to define other frames.
Using Kronecker\_tensor, basis\_selector, tensor\_selector,
contra\_index, and covar\_index, we can define tensor spaces on those
frames, assign values to tensor components, and write tensor formulas.
The formulas may be within a single frame or between frames.
Finally, with the metric tensor we can specify non-Euclidean metrics
and using the Levi-Civita symbol we can express vector cross products and
the curl operation in vector component form.
These symbols are being submitted as a content dictionary named
tensor1 to the online OpenMath repository.

\thanks{Many thanks to Weiqing Gu at Naval Research Lab for several
  conversations regarding tensors.}

\end{document}